\documentclass[]{bytedance_seed}

\usepackage{amsmath}   % (needed by many things)
\usepackage{amssymb}

\usepackage{subcaption} % loads caption compatibly
\usepackage{wrapfig}
\usepackage{booktabs}
\usepackage{makecell}

\usepackage{siunitx}
\usepackage{microtype}
\usepackage{enumitem}
\setlist{nosep,leftmargin=*,itemsep=0.25ex,topsep=0.4ex,parsep=0ex}

\usepackage{caption}
\usepackage{etoolbox}
\newcommand{\tabledefaults}{%
  \setlength{\tabcolsep}{6pt}%
  \renewcommand{\arraystretch}{1.05}%
}
  {\begingroup #1\setlength{\abovecaptionskip}{3pt}\setlength{\belowcaptionskip}{0pt}%
   \setlength{\tabcolsep}{5pt}\renewcommand{\arraystretch}{1.0}}%
  {\endgroup}
  {\begingroup #1\setlength{\abovecaptionskip}{2pt}\setlength{\belowcaptionskip}{0pt}%
   \setlength{\tabcolsep}{4pt}\renewcommand{\arraystretch}{0.97}}%
  {\endgroup}
\tabledefaults

\usepackage{algorithm}
\usepackage{algpseudocode}

\usepackage{xcolor}
\usepackage{xspace}
\usepackage{varwidth}
\usepackage{pifont}
\usepackage{mathtools}
\usepackage{amssymb}

\usepackage{CJKutf8}
\usepackage{pgfplots}
\pgfplotsset{compat=1.18}

\usepackage[toc,page,header]{appendix}
\usepackage{minitoc}

\usepackage{hyperref} % this also provides \url; no need for separate url package

\newcommand{\pluseq}{\ensuremath{\mathrel{+}=}}
\title{
Protenix-Mini+: efficient structure prediction model with scalable pairformer
}

\author[1,2,*, \dagger]{Bo Qiang}
\author[1,\dagger]{Chengyue Gong}
\author[1]{Xinshi Chen}
\author[1]{Yuxuan Zhang}
\author[1]{Wenzhi Xiao}

\affiliation[1]{ByteDance Seed}
\affiliation[2]{University of Washington}

\contribution[*]{Work done at ByteDance Seed}
\contribution[\dagger]{Equal contribution}

\abstract{
Lightweight inference is critical for biomolecular structure prediction and downstream tasks, enabling efficient real-world deployment and inference-time scaling for large-scale applications. 
While AF3 and its variants (e.g., Protenix, Chai-1) have advanced structure prediction results, they suffer from critical limitations: high inference latency and cubic time complexity with respect to token count, both of which restrict scalability for large biomolecular complexes.
To address the core challenge of balancing model efficiency and prediction accuracy, we introduce three key innovations: (1) compressing non-scalable operations to mitigate cubic time complexity, (2) removing redundant blocks across modules to reduce unnecessary overhead, and (3) adopting a few-step sampler for the atom diffusion module to accelerate inference. Building on these design principles, we develop Protenix-Mini+ — a highly lightweight and scalable variant of the Protenix model. Within an acceptable range of performance degradation, it substantially improves computational efficiency. For example, in the case of low-homology single-chain proteins, Protenix-Mini+ experiences an intra-protein LDDT drop of approximately 3\% relative to the full Protenix model—an acceptable performance trade-off given its substantially 90\%+ improved computational efficiency.

}

\date{\today}
\begin{document}
\maketitle

\newcommand{\figref}[1]{Figure~\ref{#1}}
\newcommand{\todo}[1]{\colorbox{yellow!30}{[{#1}]}}

\newcommand{\outline}[1]{%
  \colorbox{cyan!20}{%
    \begin{varwidth}{\dimexpr\linewidth-2\fboxsep\relax}
    [{\bf Outline}]  #1
    \end{varwidth}%
  }%
}

\section{Introduction}
Accurate prediction of biomolecular structures, such as proteins and nucleic acids, is the cornerstone of modern
structural biology and drug discovery, allowing insight into molecular function, interactions, and rational
design of therapeutic agents~\citep{jumper2021highly,cheng2023accurate, evans2021protein}.
Following the release of AlphaFold3~\citep{abramson2024accurate} (AF3), substantial progress has been made toward reproducing its prediction accuracy using a similar architecture composed of three main stages: 
\ding{192} atom feature embedding; \ding{193} token representation updates; and \ding{194} coordinate diffusion.
%\cy{unify diffusion and backbone name in intro and method section} 
Representative efforts include Chai-1~\citep{chai2024chai}, Protenix~\citep{team2024protenix}, Boltz-1~\citep{wohlwend2025boltz}, and RF3~\citep{corley2025accelerating}.
These developments have established a new paradigm for end-to-end biomolecular modeling, enabling predictions across proteins, RNA, DNA, and their complexes. Yet, questions remain around efficiency and generalizing to unseen biomolecules, especially for screening of \textit{de novo} designed protein~\citep{ahern2025atom,ren2025pxdesign,zambaldi2024novo}. AF3 requires $>20$ seconds for folding a biomolecular sequence with more than 1024 tokens on a NVIDIA A100 GPU, which is far from applicable in high-throughput experiments. Additionally, AF3’s time complexity is cubic with respect to token count, rendering it intractable for large structures, such as antibody-antigen complexes~\citep{bennett2024atomically}.

Lightweight inference frameworks and scalable architectures that bridge the gap between efficiency and accuracy have been widely deployed in image generation and language models. In this study, we identify several pathways to achieve both faster and scalable inference:

\textbf{Compression of $\mathcal{O}(n^3)$ operations.}
In addition to the challenges posed by deep networks, scaling current structural prediction models for large complexes is further complicated by the computationally intensive triangular attention and triangular multiplicative update operations. Although the implementation of flash attention and fused kernels~\citep{dao2022flashattention} has reduced the memory requirements to a quadratic function of the number of tokens, the number of floating point operations (FLOPs) remains cubic with respect to the length of the inputs. Our research has shown that employing linear attention mechanisms~\citep{katharopoulos2020transformers,yang2023gated} and chunking pair representations into a fixed number of patches enhances the performance of structural prediction models while increasing their scalability for large complexes.

\textbf{Efficient few-step ODE sampler.}
Recent progress in score-based generative modeling shows that the diffusion process can be approximated through ordinary differential equations (ODEs), allowing for deterministic sampling ~\citep{liu2023instaflow,lu2022dpm}. Our analysis demonstrates that models trained under different generative frameworks (such as EDM~\citep{karras2022elucidating} in AF3 and flow matching~\citep{lipman2023flowmatchinggenerativemodeling,liu2022flow,liu2023instaflow}) remain surprisingly robust even when the number of sampling steps is drastically reduced. In particular, we find that these models can still produce high-quality backbone structures with as few as two steps, and get reasonable side-chain structures with fewer than ten steps,  challenging the common assumption that high-quality samples demand dozens of iterations \citep{gong2025protenix}.

\textbf{Redundant blocks.}
%\cy{shorten}
Analyzing the open-source Protenix, a AF3-style structure predictor, we find that several blocks contribute little to final accuracy. By removing these noncritical blocks, we achieve improved efficiency with only a slight reduction in performance.

Based on these insights, we introduce \textbf{Protenix-Mini+} alongside \textbf{Protenix-Mini}, a family of lightweight Protenix models.
Across the RecentPDB and PoseBusters (PB) test sets,
Protenix-Mini exhibits only modest performance degradation relative to larger baseline models, including Protenix, Chai-1 \citep{chai2024chai}, and Boltz-1 \citep{passaro2025boltz}.
Protenix-Mini+ experiences a 1–12\% reduction in interface LDDT (iLDDT) scores across diverse test domains, but in return delivers substantially improved inference efficiency—particularly for large biomolecular complexes with long sequences.
On the AF3 antibody (AF3AB) test set, both Protenix-Mini+ and Protenix-Mini further demonstrate potentials for exploring inference-time scaling strategies, specifically, the trade-off between per-seed inference latency and the performance gains from using a larger number of sampling seeds.

\begin{figure}[ht]
    \centering
    \includegraphics[width=1.025\textwidth]{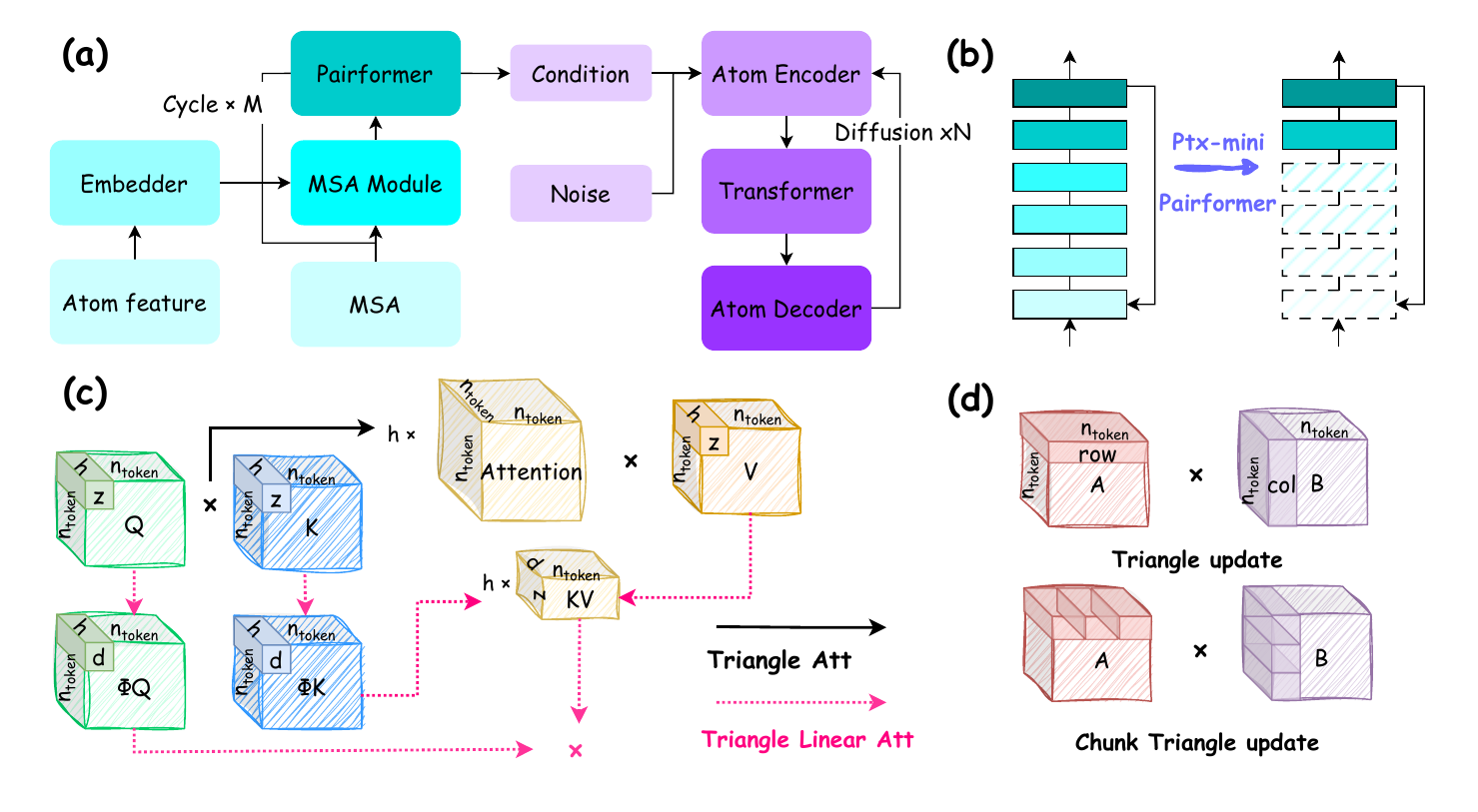}
    \vspace{-12pt}
    \caption{ \textbf{An overview of our proposed architecture.} (a) Overview of the AF3-style model architecture. The model consists of three stages: \ding{182} a feature aggregation stage that collects atomic representations; \ding{183} a token-level refinement stage that iteratively enhances representations with the deep pairformer and MSA to create a rich representation for diffusion conditioning; and \ding{184} a diffusion stage that transforms 
    randomly initialized diffusion noise into the final atomic structure.
    (b) By eliminating stacked blocks, we reduce the model size.
    (c) Triangle linear attention replaces the cubic triangle attention with a kernelized formulation that projects queries and keys into feature maps, enabling efficient updates.
    (d) Chain-aware chunking reduces the cubic cost of triangular updates by partitioning sequences into blocks and mean-pooling within each block, yielding compressed representations that enable efficient chunk triangle updates.
    }
    \label{fig:method}
\end{figure}

% \cy{todo: add performance}
% that employs a reduced number of blocks, few integration steps, and scalable pairformer operations.

%TODO: experiment summary here
\begin{table}[t]
\centering
\caption{Computational complexity of the major modules in the AF3-style architecture.}
\resizebox{\textwidth}{!}{%
\begin{tabular}{lcc}
\toprule
\textbf{Module} & \textbf{Memory Complexity} & \textbf{Theoretical FLOPs} \\
\midrule
Atom-Level embedding & $\mathcal{O}(n_{\text{atom}})$ & $3.97\text{M}~n_{\text{atom}}$ \\
Token-level refinement & $\mathcal{O}(n_{\text{token}}^2)$ & $n_{\text{cylce}} \times n_{\text{block}} \times(1.5\text{K}~n^3_{\text{token}} + 1.1\text{M}~n^2_{\text{token}}) $ \\
Diffusion Module (per step) & $\mathcal{O}(n_{\text{atom}}) + \mathcal{O}(n_{\text{token}}^2)$ & $4.9~\text{M}~n^2_{\text{token}} + 131~\text{M}~n_{\text{token}} + 9.8~\text{M}~n_{\text{atom}}$ \\
\bottomrule
\end{tabular}%
}
\label{tab:complexity}
\end{table}

\section{Preliminary: AF3 model overview and complexity analysis}
This section provides a brief overview and a computational analysis of the current AF3-style structure prediction models. We summarize the memory complexity and FLOPs in Table~\ref{tab:complexity}. The overall pipeline comprises three core stages: atom-level feature embedders, token-level representation updates, and the diffusion process, as shown in Figure~\ref{fig:method}(a).

%The model first computes the conditional signals from the sequence, multiple sequence alignment (MSA). The searched MSA results and sequence embeddings are passed through the MSA
%module and Pairformer for M cycles, yielding per-token and pairwise representations, denoted as s and z, respectively: (s, z) = Cond(Seq, MSA). These serve as conditioning inputs for the subsequent diffusion module.

Focusing on the first two stages, atomic features computed from sequence inputs are processed through the atom-level transformer and then aggregated into token representations. 
These are subsequently updated by the token-level Pairformer, which iteratively refines the pair representation using information from the MSA. 
Let Seq denote the sequence input and MSA the multiple sequence alignment. The MSA results and sequence embeddings are passed through the MSA module and Pairformer for $M$ cycles, yielding per-token and pairwise representations, denoted as $s$ and $z$ respectively:
$(s, z) = \text{Cond}(\text{Seq}, \text{MSA})$.
Because the atom transformer employs a sparse attention mechanism, its computational cost scales linearly with the number of atoms, requiring approximately $3.97\text{M} \times n_{\text{atom}}$ FLOPs. In the second stage, MSAs are integrated through a small MSA module and combined with the sequence input through the token-level Pairformer, which dominates the overall cost. Each block of the Pairformer incurs cubic complexity in the number of tokens, with theoretical FLOPs on the order of 
$n_{\text{cycle}} \times n_{\text{block}} \times (1.5\text{K}\,n_{\text{token}}^3 + 1.1\text{M}\,n_{\text{token}}^2)$.

Finally, the diffusion process iteratively moves atom coordinates from randomly initialized positions to ground-truth 3D coordinates conditioned on the pair and sequence representations from the Pairformer blocks. 
 Let $x_{1}$ denote the ground-truth structure, and $x_{0} \sim \mathcal{N}(0, \sigma_{\text{data}}^{2}\mathbf{I})$ be the initial Gaussian noise. As in standard diffusion models, structure generation follows a forward--reverse paradigm:
$
x_{1} \xrightarrow{\;\;\text{forward}\;\;} x_{0}$ and 
$x_{0} \xrightarrow{\;\;\text{reverse}\;\;} \hat{x}_{1} \approx x_{1}$
where the forward process perturbs the structure by progressively adding noise, and the reverse process attempts to recover the original signal by applying a learned denoiser. In AF3, this is implemented through a sampling algorithm, following the EDM formulation~\citep{karras2022elucidating}. At each iteration, noise $\epsilon_{t}$ is injected into the current sample $x_{t}$ according to a predefined schedule, and the diffusion module predicts a denoised version from the perturbed input. Detailed algorithm will be discussed in the Method section. %\TODO add ref
Each diffusion step costs approximately 
$
4.9\text{M}\,n_{\text{token}}^2 + 131\text{M}\,n_{\text{token}} + 9.8\text{M}\,n_{\text{atom}},
$
and with hundreds of steps, this stage contributes substantially to inference runtime. 

Altogether, while the parameters of the model remain in the scale of millions, the cubic scaling of the Pairformer and repeated diffusion steps form the primary computational bottlenecks, motivating architectural innovations such as linear attention factorization and diffusion schedule reduction.

%\begin{figure}[h]
%    \centering
%    \includegraphics[width=0.95\textwidth]{figs/architecture.pdf}
%    \caption{\textbf{Overview of the AF3-style model architecture.} The model consists of three stages: (1) an atom feature aggregation stage that collects atomic representations; (2) a token-level refinement stage that iteratively enhances these representations with the deep pairformer and MSA to create a rich representation for diffusion conditioning; and (3) a diffusion stage that transforms randomly initialized diffusion noise into the final atomic structure.}
%    \label{fig:arch_figure}
%\end{figure}

\section{Methods}
We present the key methodological innovations that underpin our approach to achieving efficient structure predictions. First, we streamline the architecture by removing redundant blocks, which results in only a minimal reduction in representational capacity as demonstrated in \ref{fig:method}(b). Second, as shown in Figure \ref{fig:method}(c) and \ref{fig:method}(d), we explore the use of linear attention and chunked triangular multiplicative updates to optimize the scalability of pairformer.
Finally, we utilize a few-step ODE to expedite the diffusion sampling process. 
% An overview of the model architecture changes are illustrated in Figure~\ref{fig:method}.

\subsection{Pruning redundant blocks}

% \begin{table}[t]
% \centering
% \begin{tabular}{l l c c c c c}
% \toprule
% \#pairformer & Mode & Prot-Prot & Lig-Prot & DNA-Prot & RNA-Prot & Intra-Prot \\
% \midrule
% 48  & -         & 0.501 & 0.650 & 0.593 & 0.363 & 0.844 \\
% 44  & Zero-shot & 0.484 & 0.631 & 0.575 & 0.344 & 0.828 \\
% 44  & Finetuned & 0.500 & 0.652 & 0.592 & 0.360 & 0.843 \\
% \bottomrule
% \end{tabular}
% \caption{The smaller model can achieve similar performance with further finetuning. We fine-tune the model with batch size 256, learning rate $10^{-3}$, and 10K iterations.}
% \label{tab:compression}
% \end{table}

The AF3 architecture comprises 48 Pairformer blocks for modeling pairwise token distances and 24 Diffusion Transformer blocks for generating 3D coordinates. While these deep networks capture complex structural dependencies, their computational cost introduces substantial overhead and raises questions about architectural redundancy---specifically, whether all layers contribute meaningfully to the final prediction. Similar observations have been reported in prior work~\citep{zhang2022all, ye2020good, lee2019wide, cheng2024survey}.
Through a search over model architectures, we found that using a 16-block Pairformer, an 8-block atom transformer, and a 1-block MSA module achieves a favorable trade-off between LDDT and efficiency.
We refer readers to the experiments for more details and comparisons.
% To investigate this, we performed block-wise ablations by progressively removing early-stage blocks in the Pairformer network. Removing the first 4 Pairformer blocks resulted in only a slight loss in structural accuracy on benchmark datasets. 
% Table~\ref{tab:compression} further shows that, with finetuning of the remaining model parameters, removing several blocks does not significantly degrade performance. This finding motivates a more compact architectural configuration.
% Building on these results, we explored two different training schemes: (1) pruning these blocks and finetuning the model, and (2) directly designing a smaller architecture configuration and training from scratch. Through a search over model architectures, we found that using a 16-block Pairformer and an 8-block atom transformer achieves a favorable trade-off between LDDT scores and efficiency.

\subsection{Scalable Pairformer}
% After reducing the number of blocks and diffusion steps, one major challenge remains: 
The primary efficiency bottlenect lies in the Pairformer’s computational complexity, which scales cubically with token count.
The cubic complexity stems from triangular attention and matrix multiplications in triangular multiplicative updates. 
Specifically, each Pairformer block comprises three key components: triangular multiplicative updates (with FLOPs  $524n_{\text{token}}^3 + 393\text{K}\,n_{\text{token}}^2$), 
triangular attention (with FLOPs $1\text{K}\,n_{\text{token}}^3 + 657\text{K}\,n_{\text{token}}^2$), 
and transition updates (with FLOPs $401\text{K}\,n_{\text{token}}^2 + 6.19\text{M}\,n_{\text{token}}$).
% The cubic bottlenecks arise from the attention mechanism in triangular attention and the matrix multiplications in triangular updates.
To mitigate this cost, we propose replacing these operations with linear attention and block-wise matrix multiplication.

\subsubsection{Triangular linear attention}\label{sec:linear_attention}
Let $i,j,k$ index tokens in the pair representation, where $i$ and $j$ denote the row and column 
indices of $z_{ij}$, and $k$ serves as a summation index over all tokens when computing attention. 
Each attention head is indexed by $h$.  
Let $\mathbf{q}_{ij}^h$, $\mathbf{k}_{kj}^h$ ($\mathbf{k}_{ik}^h$), and $\mathbf{v}_{kj}^h$ ($\mathbf{v}_{ik}^h$) 
denote the query, key, and value vectors for attention head $h$, respectively, and let $b_{ki}^h$ ($b_{jk}^h$) 
be a learned bias term. They are all computed from the given pair representation input $z_{ij}$ using a small 
linear layer or MLP. These projection layers are different for outgoing and ingoing attention, but we denote 
them the same here for simplicity.

The ingoing(outgoing) attention weights $a_{ijk}^h$ and outputs $o_{ij}^h$ can be expressed as:
\vspace{-3pt}
\begin{align*}
a_{ijk}^{h,\text{in}} = \mathrm{softmax}_k \left( \tfrac{1}{\sqrt{c}} \, 
\mathbf{q}_{ij}^h{}^\top \mathbf{k}_{kj}^h + b_{ki}^h \right),
o_{ij}^{h,\text{in}} = \mathbf{g}_{ij}^h \odot \sum_k a_{ijk}^{h,\text{in}} \mathbf{v}_{kj}^h ,
\end{align*}
\vspace{-8pt}
\begin{align*}
a_{ijk}^{h,\text{out}} = \mathrm{softmax}_k \left( \tfrac{1}{\sqrt{c}} \, 
\mathbf{q}_{ij}^h{}^\top \mathbf{k}_{ik}^h + b_{jk}^h \right),
o_{ij}^{h,\text{out}} = \mathbf{g}_{ij}^h \odot \sum_k a_{ijk}^{h,\text{out}} \mathbf{v}_{ik}^h .
\end{align*}
% \vspace{-1pt}
%\cy{do we need some figures to explain what we do? and a figure together which figure 1 to point out where is the operator}
The cubical complexity of full triangular attention comes from computing the cubic-sized attention matrices. To improve efficiency, we want to avoid directly computing the full attention matrices and reformulate the attention with a kernel function $\phi$:
\vspace{-3pt}
\begin{align*}
\hat{\mathbf{q}}_{ij}^h \coloneqq \phi\!\left(\mathbf{q}_{ij}^h + \sum_{k=1}^{n_{\text{token}}} (b_{ki}^h)^\top \right), \quad
\hat{\mathbf{k}}_{kj}^h \coloneqq \phi\!\left(\mathbf{k}_{kj}^h + \sum_{i=1}^{n_{\text{token}}} b_{ki}^h \right), \quad
\mathbf{o}_{ij}^{h,\text{in}} = 
\frac{\hat{\mathbf{q}}_{ij}^h \sum_{k=1}^{n_{\text{token}}} \left( \hat{\mathbf{k}}_{kj}^h \right)^\top \mathbf{v}_{kj}}
     {\hat{\mathbf{q}}_{ij}^h \sum_{k=1}^{n_{\text{token}}} \hat{\mathbf{k}}_{kj}^h }.
\end{align*}
\vspace{-8pt}
\begin{align*}
\hat{\mathbf{q}}_{ij}^h &\coloneqq \phi\!\left(\mathbf{q}_{ij}^h + \sum_{k=1}^{n_{\text{token}}} (b_{jk}^h)^\top \right), \quad
\hat{\mathbf{k}}_{ik}^h \coloneqq \phi\!\left(\mathbf{k}_{ik}^h + \sum_{j=1}^{n_{\text{token}}} b_{jk}^h \right), \quad
\mathbf{o}_{ij}^{h,\text{out}} = 
\frac{\hat{\mathbf{q}}_{ij}^h \sum_{k=1}^{n_{\text{token}}} \left( \hat{\mathbf{k}}_{ik}^h \right)^\top \mathbf{v}_{ik}}
     {\hat{\mathbf{q}}_{ij}^h \sum_{k=1}^{n_{\text{token}}} \hat{\mathbf{k}}_{ik}^h }.
\end{align*}
While the linear attention formulation avoids explicitly constructing cubic-sized matrices, the choice of kernel function $\phi$ remains critical for both accuracy and training stability. 
To avoid manually selecting a kernel formulation that generalizes across all Pairformer blocks, we parameterize $\phi$ as a learnable linear transform:
\begin{align*}
\phi(\bm{x}) 
&= \mathrm{concat}\!\left( \exp(\bm{W}_{\phi} \bm{x} + b_{\phi}),\; \exp(- \bm{W}_{\phi} \bm{x} - b_{\phi}) \right),
\end{align*}
where $\bm{W}_{\phi}$ and $b_{\phi}$ are learnable parameters. 
This formulation guarantees positivity while allowing block-specific adaptation of the kernel representation. 
To further accelerate inference, we implement a custom Triton kernel, as described in Appendix~\ref{appdx:flash_attention}. 
A natural question arises as to whether a single, uniform form exists for an optimal linear attention kernel. 
To investigate this, we employ a Padé approximation~\citep{baker1961pade} to identify effective kernel functions, and find that different model blocks consistently favor distinct kernel forms. 
Comprehensive experimental results are provided in Appendix~\ref{appdx:pade}.

%Related work in language modeling has similarly explored kernels that enforce positivity of attention weights, including ReLU-based activations~\citep{kasai2021finetuning} and randomized feature maps~\citep{choromanskirethinking}. 

\textbf{Distillation Loss.}
To accelerate convergence when fine-tuning from a pretrained checkpoint, we incorporate an auxiliary distillation loss during the early training stage, inspired by prior work~\citep{zhang2024hedgehog}. Specifically, we randomly select $s$ queries and compute attention maps from both the full triangular attention and our kernelized linear attention formulation. The student distributions $\tilde{a}_{ijk}^{h,\text{in}}$ and $\tilde{a}_{ijk}^{h,\text{out}}$ are defined as,
\vspace{-4pt}
\begin{equation*}
\tilde{a}_{ijk}^{h,\text{in}}
= \frac{\hat{\mathbf{q}}_{ij}^h{}^\top \hat{\mathbf{k}}_{kj}^h}
       {\sum_{m=1}^{n_{\text{token}}}\hat{\mathbf{q}}_{ij}^h{}^\top \hat{\mathbf{k}}_{mj}^h},
\qquad
\tilde{a}_{ijk}^{h,\text{out}}
= \frac{\hat{\mathbf{q}}_{ij}^h{}^\top \hat{\mathbf{k}}_{ik}^h}
       {\sum_{m=1}^{n_{\text{token}}}\hat{\mathbf{q}}_{ij}^h{}^\top \hat{\mathbf{k}}_{im}^h}.
\end{equation*}
The distillation loss is then given as the sum of cross-entropies between teacher and student,
\begin{equation*}
\mathcal{L}^{\text{distill}}
= \mathbb{E}_{i,j} \Big[
    \mathrm{CE}\!\left(a_{ij}^{h,\text{in}}, \tilde{a}_{ij}^{h,\text{in}}\right)
  + \mathrm{CE}\!\left(a_{ij}^{h,\text{out}}, \tilde{a}_{ij}^{h,\text{out}}\right)
\Big],
\end{equation*}
\vspace{-10pt}
%\cy{make a table to show the FLOPs under different tokens?}
\begin{figure}[h]
    \centering
    % Left side: text
    \begin{minipage}[t]{0.5\textwidth}
        \vspace{0pt} % force top alignment
        
where $\mathrm{CE}(p,q)$ denotes the standard cross-entropy between two categorical distributions. 

        In our experiments, we set the kernel dimension to twice the number of hidden dimensions used for the original queries and keys. Under this configuration, the FLOPs of triangular attention---considering only the attention operation itself, without the additional projections from pair representations to queries, keys, values, biases, and gates---are reduced from $524n_{\text{token}}^3$ to $51\text{K}\,n_{\text{token}}^2$. Consequently, we expect a noticeable inference speedup for biomolecules with more than $98$ tokens, assuming GPU performance is not dominated by data transfer latency. The FLOPs of replacing full attention is shown in Figure~\ref{fig:pairformer-flops}.
    \end{minipage}%
    \hfill
    % Right side: figure
\begin{minipage}[t]{0.45\textwidth}
    \vspace{0pt} % force top alignment
    \centering
    % Legend smaller
    \includegraphics[width=0.8\textwidth]{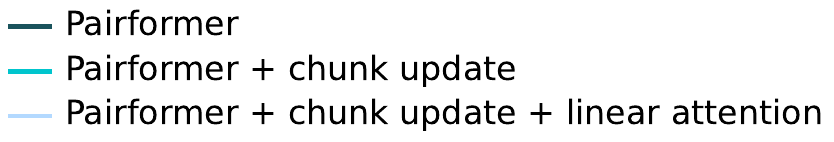}\\[-0.1em]
    % Plot immediately below, no extra padding
    \includegraphics[width=\textwidth]{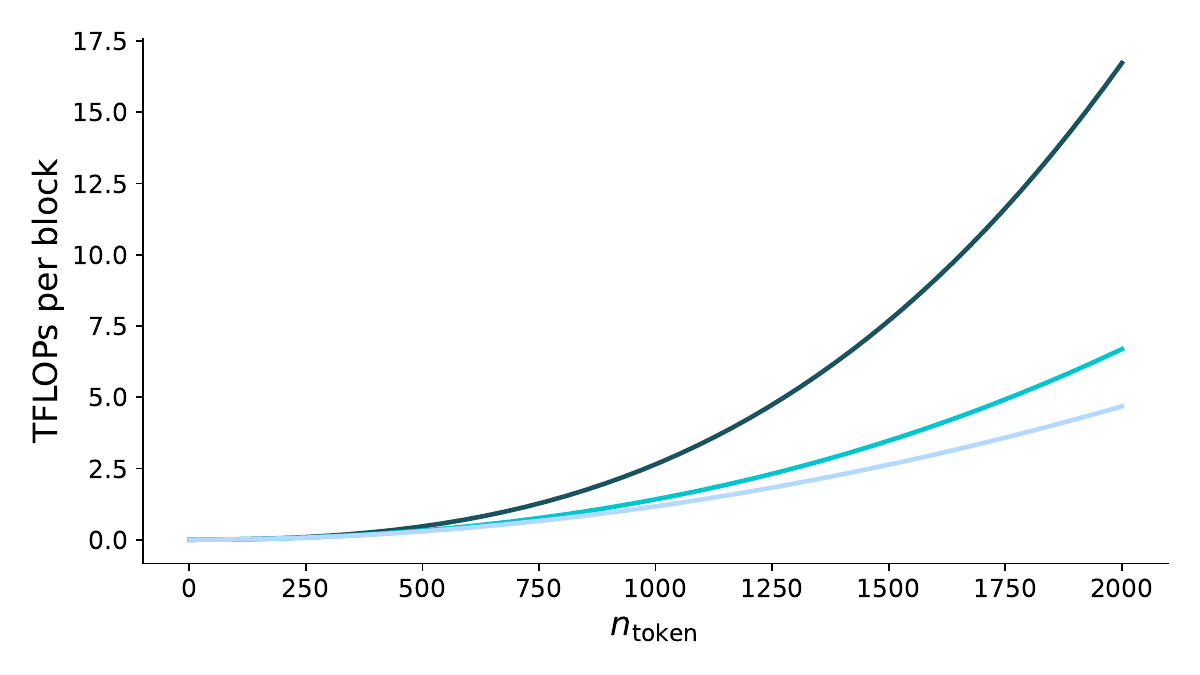}
    \vspace{-20pt}
    \caption{Comparison of computational cost (TeraFLOPs per Pairformer block) versus sequence length for different pairformer variants.}
    \label{fig:pairformer-flops}
\end{minipage}
\vspace{-10pt}
\end{figure}

% \textbf{Block-specific adaptive scheme for efficiency-accuracy trade-off.} \cy{mention in experiments}

\subsubsection{Chunking triangular update}\label{sec:trunk_update}
Denoting $\mathbf{a}_{ki}$, $\mathbf{b}_{kj}$, and $\mathbf{g}_{ij}$ as projections of the pair representation $z_{ij}$, the key update operation of triangular multiplicative updates is defined as:
\vspace{-5pt}
\begin{equation}
\tilde{z}_{ij} = \mathbf{g}_{ij} \odot 
\mathrm{LinearNoBias}\!\left(
    \mathrm{LayerNorm}\!\left(
        \sum_{k} \mathbf{a}_{ki} \odot \mathbf{b}_{kj}
    \right)
\right).
\end{equation}
\vspace{-8pt}

The cubic complexity of this module arises from the term $\sum_{k} \mathbf{a}_{ki} \odot \mathbf{b}_{kj}$. Let $\mathbf{A} \in \mathbb{R}^{n \times n \times d}$ with rows $\mathbf{A}_{k,:} = \mathbf{a}_{ki}$ and $\mathbf{B} \in \mathbb{R}^{n \times n \times d}$ with rows $\mathbf{B}_{k,:} = \mathbf{b}_{kj}$, where $d$ is the number of hidden dimensions. The computation is the mathematical the same as applying matrix multiplication $\mathbf{A}^\top \mathbf{B}$. A common strategy to accelerate this matrix operation is to employ low-rank approximations of $\mathbf{A}$ and $\mathbf{B}$. In our work, we adopt a chain-aware chunking strategy for a subset of the Pairformer blocks, striking a balance between accuracy and efficiency.

To obtain the chunk index in a chain-aware way, we proceed in three steps. First, we detect the boundaries of each chain by recording its start and end positions in the sequence. Second, we compute the length of each chain and determine how many blocks it should be divided into, based on the global sequence length and a target number of blocks, while enforcing that every chain is assigned at least one block, regardless of its length. Finally, we partition each chain into consecutive blocks of approximately equal size and assign a block index to every token. We then mean-pool token embeddings within each block, yielding compressed representations $\hat{\mathbf{A}}, \hat{\mathbf{B}} \in \mathbb{R}^{n \times r \times d}$, where $r$ denotes the number of chunks. As a result, the computational cost is reduced from $256n_{\text{token}}^3$ to $256n_{\text{token}}^2(r+1)$. 
%, where we set $d=32$.

\subsection{Few-step ODE diffusion sampling}
One of the primary bottlenecks in diffusion-based structure prediction is the substantial computational cost arising from the large number of integration steps. AF3-style models typically employ samplers with up to 200 steps. A natural question is whether such long sampling schedules are truly necessary at inference time. Somewhat surprisingly, the answer is no---provided the sampling algorithm is properly configured.

When we directly applied the default AF3 sampler with fewer inference steps, we observed severe degradation below 10 steps, often yielding broken or nonsensical structures. 
Examining the AF3 sampling Algorithm~\ref{alg:af3-sampling}, we found that, compared to the standard EDM sampler~\citep{karras2022elucidating}, AF3 increases the step size $\eta$ from $1$ to $1.5$. Although the original motivation for choosing $\eta > 1$ is not explicitly stated, prior work such as Auto Guidance~\citep{karras2024guiding} suggests that denoisers tend to underestimate velocity magnitudes in low-noise regimes ($t \to 1$, or noise level $\to 0$). While this adjustment improves stability when using a large number of steps, it becomes a key factor limiting performance in few-step settings.
% We find out that applying $\eta = 1$ for high noise-level region can make 10-step sampler get valid output complex structures.
In addition, setting $\gamma_0 = 0$ switches the sampler from an SDE, which injects noise at each diffusion step, to an ODE that traces a deterministic trajectory. This modification enables stable few-step sampling and simplifies inference by eliminating stochastic perturbations.
In practice, we notice that 2-step ODE sampler can get good backbone structures while may have weakness in generating side-chain information.
We refer the readers to our experiments for more detailed analysis.

\begin{algorithm}[t]
\caption{Sampling Algorithm in AF3}
\label{alg:af3-sampling}
\begin{algorithmic}[1]
\State \textbf{Given:} $\eta = 1.5, \; \gamma_0 = 0.8, \; \gamma_{\min} = 1, \; \lambda = 1.003, \; \sigma_{\text{data}} = 16$, and Diffuser with condition $C$.
\State \textbf{Initial:} $x_0 \sim \mathcal{N}(0, \sigma_{\text{data}} I), \; t = 0$
\State $\text{StepScheduler} \gets \text{StepSchedule}(\sigma_{\text{data}})$
\While{$t < 1$}
    \State $\Delta t \gets \text{next}(\text{StepScheduler})$
    \State $x_t \gets \text{CenterRandomAugmentation}(x_t)$ \Comment{centered and randomly rotated and translated}
    \State $\epsilon_t, \hat{t} \gets \text{NoiseSchedule}(\Delta t, t, \gamma_0, \gamma_{\min}, \mathcal{N}(0,I))$ \Comment{time-dependent noise scale}
    \State $x_t^{\text{noisy}} \gets x_t + \lambda \epsilon_t$ \Comment{add scaled noise}
    \State $x_t^{\text{denoised}} \gets \text{Diffuser}(x_t^{\text{noisy}}, \hat{t} \mid C)$ \Comment{predict denoised structure}
    \State $t \gets t + \Delta t$ \Comment{advance time}
    \State $x_t \gets x_t^{\text{noisy}} + \eta \cdot (\hat{t} - t) \cdot \text{CalVelocity}(x_t^{\text{denoised}}, x_t^{\text{noisy}}, \hat{t})$ \Comment{update via learned velocity}
\EndWhile
\end{algorithmic}
\end{algorithm}

\section{Related works}
AF3~\citep{abramson2024accurate} established a new paradigm in biomolecular structure prediction by unifying protein, nucleic acid, and ligand modeling within a single model. Follow-up works such as Chai-1~\citep{chai2024chai}, Protenix~\citep{team2024protenix}, and Boltz-1~\citep{wohlwend2025boltz} have primarily sought to replicate AF3’s architecture and training methodology to approximate its performance. Other extensions have explored novel capabilities, e.g., Boltz-2~\citep{passaro2025boltz} enabling binding affinity prediction, RF3~\citep{corley2025accelerating} incorporating chirality-aware features. Despite these advances, relatively few efforts have sought to fundamentally rethink how AF3-like models can be made more efficient.

Recent advances in generative modeling highlight two complementary directions for improving efficiency. On the sampling side, few-step diffusion methods~\citep{liu2023instaflow,lu2022dpm} accelerate inference by replacing long stochastic chains with deterministic ODE or SDE solvers~\citep{bose2023se}, adaptive noise schedules, and learned distillation strategies~\citep{song2023consistency}, substantially reducing the number of steps required without sacrificing accuracy. On the architectural side, efficient alternatives to attention in LLMs—including linear attention~\citep{katharopoulos2020transformers, shen2021efficient}, and state-space models~\citep{gu2023mamba, dao2024transformers}—offer scalable mechanisms for modeling long-range dependencies while lowering memory and FLOP costs. While these approaches have proven successful in diffusion models and LMs, they have not yet been realized in the context of AlphaFold3-style architectures. Our work is the first to adapt and demonstrate their effectiveness in this setting, enabling both faster sampling and more scalable structure prediction.
\vspace{20em}
\section{Experiments}
\subsection{Model configuration and training details}
We train a \textbf{Protenix-Mini} model comprising $16$ pairformer blocks, $8$ diffusion transformer blocks, $1$ MSA module block, $1$ atom decoder block, and $1$ atom encoder block. The model is initially trained with a batch size of $64$ for $200$K iterations using a learning rate of $10^{-3}$, following the AF3 pairformer design. All other architectural configurations match the open-sourced Protenix model~\citep{team2024protenix}\footnote{\url{https://github.com/bytedance/Protenix}}.  

After this pretraining phase, we finetune Protenix-Mini with a scalable pairformer. Specifically, all triangular attention operations are replaced with the linear attention mechanism described in Sec.~\ref{sec:linear_attention}. In addition, we apply the chunked triangular update (Sec.~\ref{sec:trunk_update}) to blocks No.~2, 3, 4, 5, 7, 8, 9, 11, 12, and 14 out of the $16$ pairformer blocks. This design is guided by our experiments, which showed that blocks 9–16 are particularly important for performance, and that leaving only the last four blocks un-chunked leads to worse performance than distributing chunking across alternating layers within the final eight. Detailed experiments are shown in Appendix~\ref{sec:chunking_pattern}.
Finetuning proceeds in three stages: first, $70$K iterations with a crop size of $384$ using the attention distillation loss; and second, $40$K iterations with a crop size of $640$ without distillation; finally, we train the confidence module upon the structure predictor. This yields the final \textbf{Protenix-Mini+} model.

% \textbf{Metrics and selectors} Model performance was assessed using interface LDDT
% (iLDDT) for multimeric interfaces, and pocket-aligned ligand RMSD for protein-ligand. For each model, we report three tiers of performance: the Oracle score, representing the best
% chain/interface structure in the generated ensemble and thereby providing an upper bound on attainable
% accuracy; the Median score, which approximates the outcome of naive random selection; and the Selected
% score, corresponding to the top-ranked structure returned by the model’s confidence score.

\textbf{Evaluation Dataset and Metrics.} For evaluation dataset, we follow the configurations in \citet{ma2025dataset}, which contains three datsets, e.g., RecentPDB, Posebusters(PB)~\citep{buttenschoen2024posebusters} and AF3 antibody dataset(AF3AB)~\footnote{\url{https://github.com/google-deepmind/alphafold3/blob/main/docs/metadata_antibody_antigen.csv}}, which contains 71 complexes and 166 interfaces. 
Both RecentPDB and AF3AB are filtered to get low-homology sequences from the training set.
For AF3AB, we infer the model with seeds ranging from 1 to 50 to investigate the inference-time scaling tradeoff—whether few-step ODE samplers can maintain comparable performance as sample size scale.
To fairly compare the structure prediction performance, we report the median performance among 5 random seeds in the this paper. 

\textbf{Inference Configuration.}
During inference, we use $4$ cycles for the backbone model and $2$-step ODE sampler for the diffusion model in Alg.~\ref{alg:af3-sampling} to save computation.
For RecentPDB and PB, we compare the difference between $4$ and $10$ cycles. Previous work has ablation the effect of number of diffusion step, and we refer the readers to \citet{gong2025protenix} for the diffusion step ablation.
For AF3AB, a much more difficult test set, we do ablation to demonstrate the impact of $n_\text{cycle}$.

\begin{table}[t]
\centering
\caption{Metrics on test datasets. On RecentPDB, we report interface LDDT and intra protein median results. On PB, we report the rate for ligand RMSD $< 2$ as `Ligand SR' (SR=success rate). We denote Protenix with scalable blocks as Protenix+. Protenix-Tiny refers to Protenix-Mini with further reduced 8 Pairformer blocks. }
\label{tab:recentpdb}
\scalebox{0.92}{
\begin{tabular}{l|cc|cccccc}
\toprule
\textbf{Model}& $n_\text{cycle}$ & Diff Step & Intra-Prot & Prot-Prot & Protein-Lig & Prot-RNA & Prot-DNA & Ligand-SR\\
\midrule
Chai-1 & 10 & 200 & 0.837 & 0.500 & 0.444 & 0.153 & 0.258 & 0.692 \\
Boltz-1 & 10 & 200 & 0.840 & 0.500 & 0.493 & 0.176 & 0.306 & 0.724 \\
Protenix & 10 & 200 & 0.830 & 0.516 & 0.510 & 0.178 & 0.281 & 0.795 \\
\midrule
Protenix+ & 4 & 2 & 0.815 & 0.454 & 0.465 & 0.144 & 0.296 & 0.749 \\
Protenix-Mini & 4 & 2 & 0.828 & 0.465 & 0.493 & 0.139 & 0.270 & 0.727 \\
Protenix-Mini+& 4 & 2 & 0.797 & 0.404 & 0.448 & 0.138 & 0.272 & 0.705 \\
Protenix-Tiny & 4 & 2 & 0.766 & 0.369 & 0.422 & 0.112 & 0.240 & 0.675 \\
\midrule
Protenix-Mini+& 10 & 2 & 0.801 & 0.411 & 0.451 & 0.138 & 0.275 & 0.711 \\
\bottomrule
\end{tabular}}
\end{table}

\subsection{Experimental results for structure prediction}
\textbf{Main  Results}
We demonstrate the main results on RecentPDB and PB in Table \ref{tab:recentpdb}.
We observe that adopting either efficient operations (e.g., linear attention, chunking) or scaling down model size individually results in only a slight performance drop across diverse domains. Specifically,
\ding{182}
For protein-ligand, RNA-protein, DNA-protein interfaces, and PB benchmarks, Protenix-Mini and Proteinx-Mini+ get comparable performance with Chai-1 and Boltz-1.
\ding{183}
Critically, combining these two strategies (i.e., Protenix-Mini+) delivers a more favorable efficiency-performance trade-off in efficiency-constrained settings, outperforming naive tiny model configurations (e.g., Protenix-Tiny) alone.
For intra protein and protein protein inference LDDT metrics, Protenix-Mini+ is better than Proteinx-Tiny ($0.797$ v.s. $0.766$ for intra protein LDDT and $0.404$ v.s. $0.369$ for protein protein interface LDDT).
\ding{184}
Increasing the number of inference cycles slightly improves performance—this finding motivates further inference AF3AB dataset, where cycle adjustments may enhance accuracy for antibody-antigen complexes.
\ding{185}
On the RecentPDB dataset, both Protenix-Mini and Protenix+ exhibit only slight performance drops across different interface types. However, further reducing model size (e.g., Protenix-Tiny) or over-applying efficiency optimizations leads to more significant degradation in protein protein interface and intra protein metrics (Protenix-Mini+ and Protenix-Tiny).
\ding{186}
In practice, when the target complex comprises long sequences (where inference efficiency is critical), we recommend Protenix-Mini+ to prioritize computational efficiency without excessive accuracy trade-offs. Conversely, when users prioritize high-quality structure prediction over efficiency (e.g., for scenarios requiring precise side-chain modeling), Protenix-Mini or larger baseline models (e.g., Chai-1, Protenix, AF3) are the preferred choices.
% On PB, we notice that the sampler plays the important role.
% We demonstrate this more in details in the following.

\textbf{AF3AB Results.}
On AF3AB, a much harder testing benchmark, we find out that the number of cycles and the diffusion sampler play a more important role for the performance. 
We summarize the results in Table \ref{tab:af3ab}.
\ding{182}
We notice that both Protenix-Mini and Protenix-Mini+ get slightly worse performance when $n_\text{cycle} = 10$. Reducing the number of cycles to $4$ get much worse performance. 
\ding{183}
2-step ODE sampler can generate enough good backbone structure compared to the 200-step AF3 sampler. 
\ding{184}
As displayed in Table \ref{tab:af3ab-seed}, both Protenix+ and Protenix-Mini+ can get better structures when we increase the sample size. 5-seed Protenix+ can yield similar results as 1-seed Protenix. We leave efficient inference scaling for future work.
% 10-step ODE sampler get slightly better results, and we recommend our 10-step sampler when a high-quality side-chain or ligand pocket structure is required.
% \ding{185}
% When scaling the inference with number of samples and number of cycles, we notice that 

\begin{table}[t]
\centering
\caption{Metrics on AF3AB. We report the median metrics over 5 different random seed, where each seed have 5 random noise. `DockQ SR' refers to the rate of DockQ $> 0.23$, while `iLDDT' represents interface LDDT.}
\label{tab:af3ab}
\scalebox{.94}{
\begin{tabular}{l|l|c|cc}
\toprule
\textbf{Model} & Sampler & $n_\text{cycle}$ &  DockQ SR & iLDDT \\
\midrule
Protenix       & 200-step AF3 & 10 & 0.185 & 0.169 \\
\hline
Protenix-Mini  & 200-step AF3 & 10 & 0.163 & 0.146\\
Protenix-Mini  & 2-step ODE & 10 & 0.163 & 0.143 \\
Protenix-Mini  & 2-step ODE & 4 & 0.141 & 0.114 \\
\hline
% Protenix-Mini+  & 200-step AF3 & 10 & 0.127 & 0.124 \\
% Protenix-Mini+  & 10-step ODE & 10 & & \\
Protenix-Mini+  & 2-step ODE & 10 & 0.127 & 0.122 \\
Protenix-Mini+  & 2-step ODE & 4 & 0.096 & 0.109 \\
\bottomrule
\end{tabular}}
\end{table}

\begin{table}[t]
\centering
\caption{Metrics on AF3AB. For Protenix-Mini+ and Protenix+, we report the impact for number of seed when we apply ipTM score as a selector. Here, we report the selected iLDDT.}
\label{tab:af3ab-seed}
\scalebox{.94}{
\begin{tabular}{l|l|c|ccc}
\toprule
\textbf{Model} & Sampler & $n_\text{cycle}$ & 1 & 5 & 50 \\
\midrule
Protenix & 200-step AF3 & 10 & 0.169 & 0.192 & 0.231 \\
\hline
Protenix+ & 2-step ODE & 10 & 0.122 & 0.162 & 0.218 \\
Protenix-Mini+  & 2-step ODE & 10 & 0.109 & 0.152 & 0.183  \\
\bottomrule
\end{tabular}}
\end{table}

\textbf{Ablation Study.}
We conduct ablation studies on several training strategies, including removing the distillation loss, fine-tuning versus training from scratch, using only linear attention, applying only chunked updates, and chunked updates without block selection. In each ablation setting, we followed the training configuration in stage-1 pretraining with a crop size of 384.

We evaluate these models on a RecentPDB subset containing 384 complexes, and the LDDT results are reported in Table~\ref{tab:mini-ablation}. The performance differences between Protenix-Mini+ trained with and without distillation, or when initialized from scratch versus fine-tuning from Protenix-Mini checkpoints, are relatively small. Nevertheless, both distillation and fine-tuning accelerate convergence. By contrast, performance drops sharply when chunked triangular updates are applied to all blocks, demonstrating that block selection is critical when adopting chunked updates. Finally, when comparing the linear-attention-only and chunked-update variants, we observe that Prot–Prot and Prot–Lig tasks suffer the most performance degradation when applying both operations..

\begin{table}[t]
\centering
\caption{Performance comparison of \textsc{mini+} variants under different ablations. Median values of intra-protein LDDT and interface LDDT are reported for a RecentPDB subset with 384 data instances. We do not report Prot-RNA and Prot-DNA interface prediction results, given that the insufficient data size in this subset may lead to spurious (i.e., random) outcomes.}
\label{tab:mini-ablation}
%\resizebox{\textwidth}{!}{%
\scalebox{1.}{
\begin{tabular}{lccccc}
\toprule
\textbf{Model Variant} & Complex & Intra-Prot & Prot-Prot & Prot-Lig & PoseBuster \\
\midrule
Protenix-Mini+          & 0.796 & 0.820 & 0.423 & 0.596 & 0.700 \\
\midrule
w/o distill             & 0.796 & 0.819 & 0.418 & 0.616 & 0.698 \\
Only linear attention   & 0.800 & 0.829 & 0.438 & 0.624 & 0.712 \\
Only chunked update     & 0.799 & 0.830 & 0.440 & 0.611 & 0.691 \\
All blocks with chunked update & 0.744 & 0.786 & 0.354 & 0.557 & 0.552 \\
Mini+ from scratch      & 0.789 & 0.818 & 0.424 & 0.604 & 0.692 \\
\bottomrule
\end{tabular}}
\end{table}

\begin{figure}[t]
    \centering
    \includegraphics[width=0.98\textwidth]{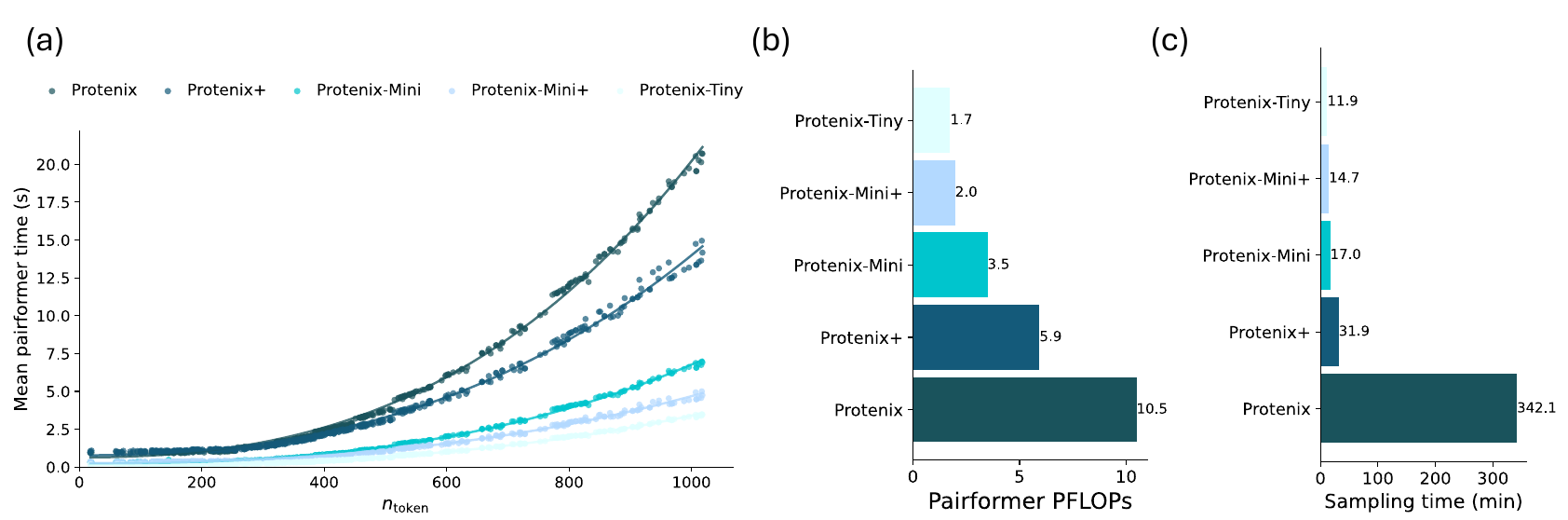}
    \caption{Runtime and efficiency comparison across Protenix model variants.
(a) Pairformer module runtime on GPUs with compute capability comparable to NVIDIA H100. Protenix scales cubically with sequence length, while lighter variants achieve reduced inference time.
(b) Pairformer computational cost in PFLOPs.
(c) Full model sampling time (minutes), showing substantial speedups for Mini, Mini+, and Tiny models from fewer blocks, scalable operations, and fewer diffusion steps.}
    \label{fig:efficiency_sum}
\end{figure}

\subsection{Efficiency analysis}
To assess efficiency, we benchmark the inference runtime of Protenix, Protenix+, Protenix-Mini, and Protenix-Mini+ on 384 sampled complexes from the recent PDB test set using GPUs with compute capability comparable to NVIDIA H100. The running time of the Pairformer components of each model variant is plotted in Figure~\ref{fig:efficiency_sum}. 

From Figure~\ref{fig:efficiency_sum}(a), we observe a consistent trend when adding scalable operations to Protenix and Protenix-Mini. Both Protenix+ and Protenix-Mini+ scale more favorably with increasing sequence length. As shown in Figure~\ref{fig:efficiency_sum}(a,b), the FLOPs and runtime gaps between Protenix-Mini+ and Protenix-Tiny are relatively small, while Protenix-Mini+ achieves higher accuracy. This indicates that, for Mini models, incorporating scalable operations yields a better performance–efficiency trade-off than further pruning deep networks.

Examining the full inference time in Figure~\ref{fig:efficiency_sum}(c), we find that introducing a few-step ODE sampler substantially reduces the overall computational cost. Besides, Scalable operations alone lower the runtime from 17 minutes to 14.7 minutes on this subset (average length 445 tokens). Based on the fact that we have removed most cubic scaling terms, the performance gain is expected to be even larger for longer sequences.

\section{Conclusion}
In this work, we represent Protenix-Mini+, which bridges the gap between high-accuracy structure prediction and practical deployment needs. 
By enabling fast inference for large biomolecular complexes and reducing computational barriers, it paves the way for broader applications in structural biology—from high-throughput drug discovery to the characterization of large multi-molecular assemblies-both scenarios in which computational efficiency remains one of the critical bottlenecks to practical deployment.
% Besides antibodies, we sparsely discussed other specific domains of biomolecular complexes modeling. 
% Improving side-chain quality and confidence relation with experimental binding affinity will be the focus of future work.
One ongoing direction is to enhance the performance of Protenix-Mini+ for specific high-impact domains—most notably antibody-antigen complex prediction.
A second future direction focuses on exploring efficient inference-time scaling strategies—particularly for multi-seed sampling, and further distill the capabilities of inference-time extensibility into the backbone model.
A third avenue involves pursuing joint development and optimization in collaboration with hardware platforms (e.g., GPU/TPU accelerators) and optimize the kernel. By co-designing model architectures with hardware-specific optimizations, we target at real-time acceleration.

\section*{Acknowledgement}
We thank Yuxuan Song and other team members for their insightful discussions.

\bibliography{example_paper}
\bibliographystyle{icml2025}

%%%%%%%%%%%%%%%%%%%%%%%%%%%%%%%%%%%%%%%%%%%%%%%%%%%%%%%%%%%%%%%%%%%%%%%%%%%%%%%
%%%%%%%%%%%%%%%%%%%%%%%%%%%%%%%%%%%%%%%%%%%%%%%%%%%%%%%%%%%%%%%%%%%%%%%%%%%%%%%
% APPENDIX
%%%%%%%%%%%%%%%%%%%%%%%%%%%%%%%%%%%%%%%%%%%%%%%%%%%%%%%%%%%%%%%%%%%%%%%%%%%%%%%
%%%%%%%%%%%%%%%%%%%%%%%%%%%%%%%%%%%%%%%%%%%%%%%%%%%%%%%%%%%%%%%%%%%%%%%%%%%%%%%
\newpage
\appendix
\onecolumn
\section{Supplement experiments}
\subsection{Kernel searching for linear attention}\label{appdx:pade}
\begin{figure}[ht]
    \centering
    \begin{tabular}{c}
        \includegraphics[width=0.91\textwidth]{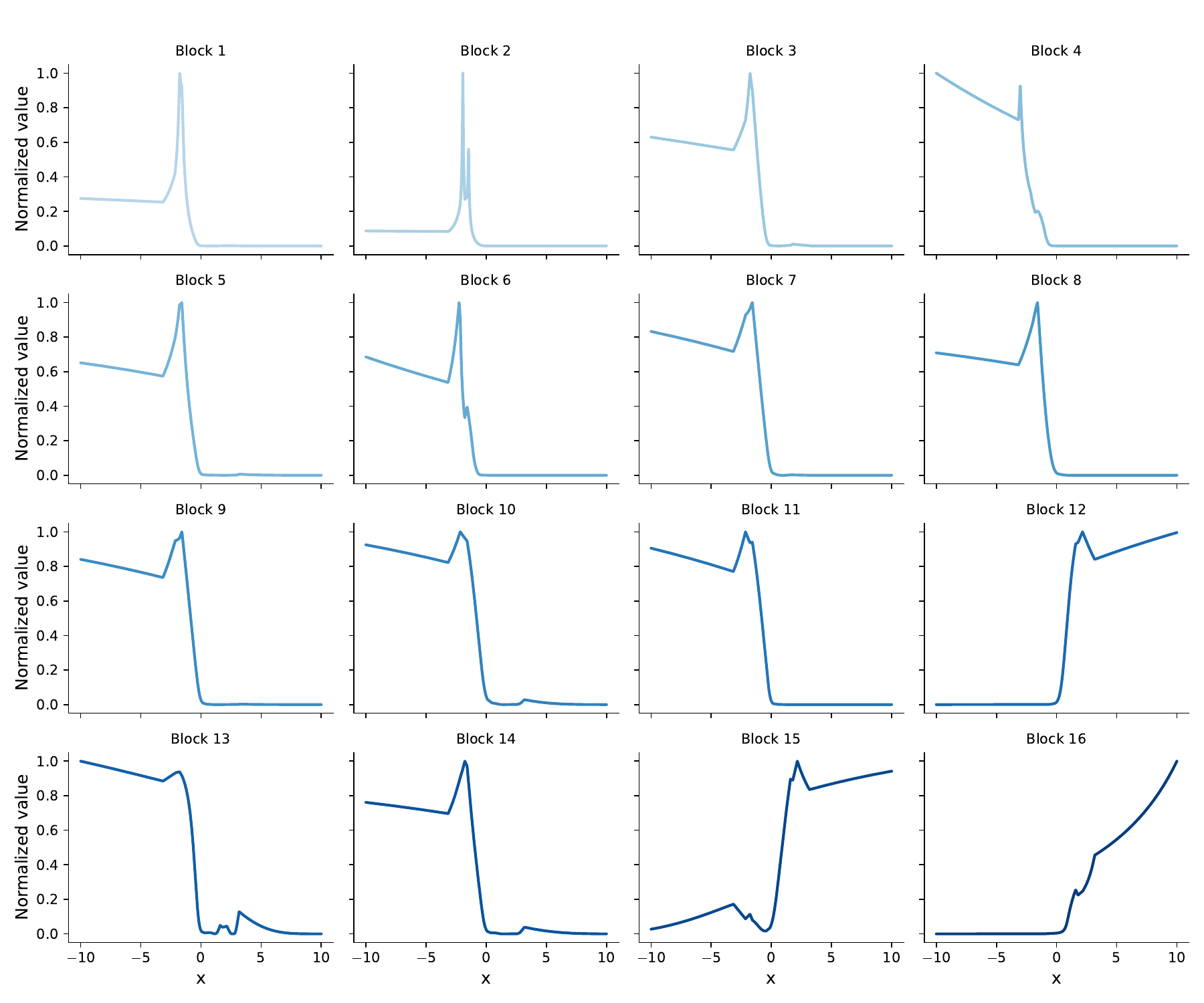}
    \end{tabular}
    \caption{Normalized Pade approximation across 16 pairformer blocks.}
    \label{fig:pade_approx}
\end{figure}

As discussed in \ref{sec:linear_attention}, the key for designing efficient linear attention is to choose the kernel. 
Before diving into using a non-negative learnable layer for the attention kernel, we carry out a kernel function search that is inspired by the pad{\'e} approximant~\citep{baker1961pade}, which has the following form:
\[
F(x) = \frac{P(x)}{Q(x)}
= \frac{\sum_{j=0}^{m} a_j x^j}{1 + \left| \sum_{k=1}^{n} b_k x^k \right|}
= \frac{a_0 + a_1 x + a_2 x^2 + \cdots + a_m x^m}{1 + \left| b_1 x + b_2 x^2 + \cdots + b_n x^n \right|}.
\]
This formulation has proven to be an effective approximation to popular non-linear activation functions~\citep{molina2019pad}. To enforce non-negativity, we use $\phi(x) = \text{SoftPlus}(F(x))$ as kernels and train only on the distillation loss starting from a pretrained Protenix-mini checkpoints. We plot the learnt kernel functions in Figure~\ref{fig:pade_approx}. This clearly shows the variability in the functional forms of different pairformer blocks. Earlier blocks tend to produce a near Dirac delta function, while later blocks show a "ReLU-like" form that only one half of the range has signals. The positive and negative properties suddenly flip at block 12 and flip back at block 13. Consequently, it is not sufficient to impose a single, uniform kernel across all blocks.

\subsection{Evaluation of chunking patterns}\label{sec:chunking_pattern}

\begin{table}[t]
\centering
\caption{Performance for applying chunking triangular update across different block subsets. 
Numbers in parentheses indicate the number of selected blocks.}
\label{tab:block-subsets}
%\resizebox{\textwidth}{!}{%
\begin{tabular}{lccc}
\toprule
\textbf{Blocks} & {Intra Protein} & {Prot-Prot} & {Protein-Lig} \\
\midrule
1,2,3,4,5,6,7,8 (8)              & 0.810 & 0.405 & 0.579  \\
2,4,6,8,10,12,14,16 (8)          & 0.809 & 0.408 & 0.575  \\
9,10,11,12,13,14,15,16 (8)       & 0.792 & 0.374 & 0.545  \\
1,2,3,4,5,6,7,8,9,10 (10)        & 0.796 & 0.404 & 0.573  \\
1,2,3,4,5,6,7,8,10,12,14,16 (12) & 0.808 & 0.389 & 0.575  \\
1,2,3,4,5,6,7,8,9,10,11,12 (12)  & 0.807 & 0.381 & 0.553  \\
\bottomrule
\end{tabular}%
%}
\end{table}

In our mini+ model, we applied chunking triangular updates to only the specific 10 blocks (No.2, 3, 4, 5,
7, 8, 9, 11, 12, and 14) among all 16 pairformer blocks. 
This choice is motivated by our small-scale experiments, where we finetuned the Protenix-Mini model by applying chunking triangular updates across different subsets of pairformer blocks. 
All experiments are running with a batch size of 16 and 250k steps of learning rate $0.0018$ without warmup. The results are shown in Table~\ref{tab:block-subsets}. 
Comparing chunking the first 8 blocks with the last 8 blocks, we observe that chunking early blocks (1–8) preserves strong intra-protein accuracy (0.810) and competitive ligand performance (0.579), whereas applying chunking only to the later blocks (9–16) significantly reduces overall accuracy, especially for ligand (0.545) and DNA binding (0.382). Distributing chunking across both early and mid-stage blocks (e.g., 2,4,6,8,10,12,14,16) leads to a better trade-off, improving Prot-DNA (0.489) and Prot-RNA (0.199) while maintaining high intra-protein performance. 
In addition, increasing the number of chunked blocks (10–12) hurt the performance of protein-protein interfaces and intra-protein lddt. 

Based on these findings, we select a set of 10 blocks spanning early and mid-stage layers while avoiding late blocks that are critical for final refinement.

\section{Supplement methods}
\subsection{Tiled triangular linear attention kernel}\label{appdx:flash_attention}
Triangular attention has been a core component of structure prediction models since AlphaFold2~\citep{jumper2021highly}. Several works have introduced efficient CUDA kernels to accelerate this inherently cubic operation, \textit{e.g.}, CuEquivariance~\citep{nvidia_cuequivariance_doc} and DeepSpeed\_Evo~\citep{deepspeed_ds4sci_evoformerattention}. To further improve the performance of our triangular linear attention, and to enable fair comparison with these CUDA-optimized baselines, we implemented a tiled triangular linear attention kernel in Triton~\citep{tillet2019triton}.

In the forward pass, queries and keys are projected blockwise into positive and negative feature maps, and the Triton kernel accumulates partial sums of key statistics and value products across sequence tiles. This tiling strategy avoids materializing large intermediate tensors and allows reuse of keys and values in on-chip memory. Numerators and denominators are then assembled from the accumulated terms, yielding outputs identical to the manual formulation but with lower memory footprint and improved cache locality. In practice, this kernel achieves faster execution on A100 GPUs when the Triton kernels are precompiled. A detailed pseudocode for the linear attention kernel is provided in Algorithm~\ref{alg:tiled_linear_attn}.

\begin{algorithm}[t]
\caption{Tiled Linear Attention}
\label{alg:tiled_linear_attn}
\begin{algorithmic}[1]
\Require Queries $X_q \in \mathbb{R}^{B \times H \times N \times F}$, 
         Keys $X_k \in \mathbb{R}^{B \times H \times T \times F}$, 
         Values $V \in \mathbb{R}^{B \times H \times T \times M}$
\Require Projection weights $W_q, W_k$, biases $b_q, b_k$, numerical $\varepsilon$
\Ensure Output $O \in \mathbb{R}^{B \times H \times N \times M}$
\Statex

\For{each query block $Q[n]$ of size $\text{BLOCK}_N \times F$}
    \State Project queries: $Q_{\text{blk}} = X_q[n] W_q + b_q + q_\text{bias}[n]$
    \State Accumulators: $KV_{\text{pos}}, KV_{\text{neg}}, K_{\text{sum,pos}}, K_{\text{sum,neg}} \gets 0$
    \For{each key block $K[t]$ of size $\text{BLOCK}_T \times F$}
        \State Project keys: $K_{\text{blk}} = X_k[t] W_k + b_k + k_\text{bias}[t]$
        \State Compute feature maps: $K_{\text{pos}} = e^{K_{\text{blk}}},\; K_{\text{neg}} = e^{-K_{\text{blk}}+2k_\text{bias}}$
        \State Update accumulators:
        \[
            KV_{\text{pos}} \pluseq K_{\text{pos}}^\top V[t], \quad
            KV_{\text{neg}} \pluseq K_{\text{neg}}^\top V[t]
        \]
        \[
            K_{\text{sum,pos}} \pluseq \text{sum}(K_{\text{pos}}), \quad
            K_{\text{sum,neg}} \pluseq \text{sum}(K_{\text{neg}})
        \]
    \EndFor
    \State Compute query maps: $Q_{\text{pos}} = e^{Q_{\text{blk}}},\; Q_{\text{neg}} = e^{-Q_{\text{blk}}+2q_\text{bias}}$
    \State Numerator: $\text{Num} = Q_{\text{pos}} KV_{\text{pos}} + Q_{\text{neg}} KV_{\text{neg}}$
    \State Denominator: $\text{Den} = Q_{\text{pos}} K_{\text{sum,pos}} + Q_{\text{neg}} K_{\text{sum,neg}} + \varepsilon$
    \State Output: $O[n] = \text{Num} \oslash \text{Den}$
\EndFor
\end{algorithmic}
\end{algorithm}

\end{document}